\documentclass{iopart}
\usepackage{graphicx}
\usepackage{epstopdf}

\usepackage{iopams}
\begin{document}

\title{Quasinormal modes for asymptotic safe black holes}

\author{Dao-Jun Liu, Bin Yang, Yong-Jia Zhai, Xin-Zhou Li}
\address{Center for Astrophysics, Shanghai Normal University, 100 Guilin Road, Shanghai 200234,China}
\ead{djliu@shnu.edu.cn}

\begin{abstract}
Under the hypothesis of asymptotic safety of gravity, the static, spherically symmetric black hole solutions in the infrared  limit are corrected by non-perturbative effects. Specifically, the metric is modified by the running of gravitational couplings.  In this work, we investigate the effects of this correction to the quasinormal modes (QNMs) of a test scalar field propagating in this kind of black hole background analytically and numerically.  It is found that although the quasi-period frequencies and the damping of oscillations are respectively enhanced and weakened by the quantum correction term, the stability of the black hole remains.
\end{abstract}

\maketitle

\section{Introduction}
\label{sec:introduction}
Constructing a consistent ultraviolet (UV) complete theory of gravity is one of the diligently pursuing goals of theoretical physicists today.
It is well-known that the usual quantization procedure for general relativity (GR) leads to a non-renormalizable quantum field theory, so we have to fix an infinite number of terms to renormalize the perturbation theory, which indicates that GR is perturbatively nonrenormalizable. It has been pointed out for a long time that if higher derivative terms is included in the action of the gravity theory, a perturbatively renormalizable theory may be obtained due to higher derivative propagators that soften the divergence of the perturbative quantization \cite{Stelle1977}.   However, these higher derivative terms will generically  introduce ghosts which will result in instability of the theory.
  Fortunately, it is still possible that gravity constitutes a renormalizable field theory at the nonperturbative level. In the so-called asymptotically safe scenario\cite{Weinberg1979}, the key ingredient is a non-Gaussian fixed point (NGFP) of the gravitational renormalization group (RG) flow which controls the behavior of the theory at very high energies  and ensures the absence of unphysical UV divergences. Provided that the NGFP has a finite number of unstable directions, the resulting fundamental theory  should be a perturbatively nonrenomalizable theory.  This kind of theory has been extensively studied, see  \cite{Niedermaier2006,Niedermaier2007} for a review. There are evidences that the ghosts in higher dervative theories may be removed  when the UV limit of gravity is restricted by a fixed point in the asympototically safe (AS) scenario \cite{Benedetti2009}.

  It is believed that black hole physics provides a window into the quantum nature of gravity.  It is interesting to understand how the theory modifies the conventional black hole solution by taking into account the quantum corrections naturally incorporated into the modified gravity thoeries.
Under the assumption that the leading order quantum corrections to the black hole spacetime are captured by running the Newtonian constant $G_N$, some authors have investigated the black hole solutions in AS  gravity scenario\cite{Bonanno2000,Fall2010}.  Recently, Cai and Easson \cite{Cai2010} developed an effective method of finding vacuum solutions to Einstein's equation derived from the AS gravity with higher derivative terms and present an exact form of a Schwarzschild-(anti)-de-Sitter solutions with running gravitational coupling parameters.

Perturbations of black holes have been intensively studied in the past few decades with relation to black hole stability, gravitational wave detection  and gauge/gravity dualities (for  recent comprehensive reviews, see Refs.\cite{QNM2009, QNM2011} and references therein).   Quasinormal modes(QNMs) is the quasinormal ringing  of the background spacetime under perturbations, which  are believed to be the characteristic sound of  black holes. QNMs would lead to the direct identification of the black hole existence through the imminent gravitational wave observation.

It is also believed that the quantum-gravity corrections to astrophysical black holes should be utterly undetectable, however, for miniature black holes such as those might be produced in the high energy experiments in the Large
Hadron Collider at CERN, the quantum effects may be significant. The study of QNMs for mini black holes has its own motivations from theoretical aspects. As a matter of fact, in the past few years many investigations on the QNMs have been performed  for these mini black holes, most of which are done in the scenarios with extra dimensions(see for example, \cite{Cardoso2003a,Abdalla2007}) and some of which are studied in the  models inspired by string and other fundamental theories(see for example, \cite{Li2001,Konoplya2008,Ishihara2008,He2008}).

In this work, we would like to consider the behavior of a test scalar field propagating in an AS black hole background.
The paper is organized as follows: 
we first give a brief review on the static and spherically symmetric black hole solution in the asymptotic safe gravity.
Then,   the wave equation that govern the quasinormal ringing of black hole in the IR limit of asymptotic safe quantum gravity is derived and  QNMs of scalar perturbations are calculated analytically and numerically. 
Finally, 
we summarize our results and give some discussions.

\section{ Black hole solution in asymptotic safe gravity}
\label{sec:BH}
We start with a generally covariant high derivative gravitational theory with effective action involving a momentum cufoff $p$ \cite{Cai2010},
\begin{eqnarray}\label{eq:action-1}
\Gamma_p[g_{\mu\nu}]&=&\int d^4x\sqrt{-g}\left[p^4g_0(p)+p^2g_1(p)R\right.\nonumber\\
&& +g_{2a}(p)R^2+g_{2b}(p)R_{\mu\nu}R^{\mu\nu}+g_{2c}(p)R_{\mu\nu\sigma\rho}R^{\mu\nu\sigma\rho}\nonumber\\
&&\left. +\mathit{O}(p^{-2}R^3)\right],
\end{eqnarray}
where $g$ is the determinant of the metric tensor $g_{\mu\nu}$, $R$  is the Ricci scalar, $R_{\mu\nu}$ is the Ricci tensor and  $R_{\mu\nu\sigma\rho}$ is the Riemann tensor. The coefficients $g_i(i = 0, 1, 2a, 2b, \cdots)$ are dimensionless coupling parameters and are functions of the UV cutoff $p$. The functional gravitational RG equations are based on a momentum cutoff for the propagating degrees fo freedom and capture the nonperturbative information about the gravitational theory.
The couplings satisfy the following equations,
\begin{equation}
\frac{d g_i(p)}{d\ln p}=\beta_i[g(p)],
\end{equation}
where the beta functions can be derived  from the gravitational renormalization group equation which is dependent upon  suitable infrared cutoffs which suppress the propagation of field modes with momenta below a momentum scale \cite{Codello2006}.
By taking into account the action (\ref{eq:action-1}) together with the gauge-fixing and ghost terms and up to the one-loop approximation,
the detailed form of the  beta functions for the couplings can be obtained.
The conditions for asymptotic safety require that all the functions $\beta_i=0$ when the coupling paramters $g_i$ approach a fixed point $g_i^\ast$. By  solving the beta functions,  it is observed that there exist a Gaussian fixed point and a non-Gaussian fixed point (see Refs.\cite{Niedermaier2006,Codello2006} for the detail calculation).

For the action containing up to four derivative of the metric, the central results \cite{Cai2010} for the gravitational coupling and cosmological constant are
\begin{eqnarray}
p^4g_0&\simeq& -\frac{\Lambda(1+{\eta} p^2 G_N)(1+{\xi} p^2G_N)}{8\pi G_N},\\
p^2g_1&\simeq&\frac{1+{\xi} p^2 G_N}{16\pi G_N},\label{eq:g_1}
\end{eqnarray}
where $G_N$ and $\Lambda$ are the values of the gravitational coupling and the cosmological constant in the IR limit($p\rightarrow 0$) which should be determined by observations. The value of $\eta$ and $\xi$ are determined by the values of the running coupling parameters $\lambda(p)g_{2a}$, $\lambda(p)g_{2b}$ and $\lambda(p)g_{2c}$ at the nontrivial fixed point.
Here the coefficient $\lambda$ has the logarithmic form which approaches asymptotic freedom
\begin{equation}
  \lambda(p)=\frac{\lambda_0}{1+\frac{133}{10(4\pi)^2}\lambda_0\ln(p/M_p)},
\end{equation}
where $\lambda_0$ is the value of the coefficient $\lambda$ at the Planck scale. According to the explicit form of beta functions presented in Ref.\cite{Codello2006}, it is not difficult to obtain that 
$\xi\approx 0.72$ and $\eta \approx 0.22$. Obviously, the parameters $\xi$ and $\eta$ represent the high energy quantum corrections to classic Newton's constant $G_N$ and cosmological constant $\Lambda$. Note that it is reasonable to imagine that different choices of infrared cutoff functions will lead to different values of $\xi$ and $\eta$, however, the detailed calculations \cite{Codello2009} show that the difference among the values of $\xi$ and $\eta$  obtained from different cutoff functions is not very large.

To obtain a black hole solution, let us assume a static, spherically symmetric metric, which can be written as
\begin{equation}\label{eq:metric}
ds^2=-f(r)dt^2+\frac{dr^2}{f(r)}+r^2d\Omega_2^2,
\end{equation}
by choosing the Schwarzschild gauge  which is ensured by the Cauchy theorem for a Riemannian geometry with unique boundary.

Substituting the above metric into the generalized vacuum  Einstein field equations,
\begin{equation}\label{eq:Einstein}
\tilde{G}^{\mu\nu}\equiv \frac{\delta\Gamma_p[g_\mu\nu]}{\delta g_{\mu\nu}}=0,
 \end{equation}
 and taking into account the consistency relation with infrared limit solution, it is derived in Ref.\cite{Cai2010} that
\begin{equation}\label{eq:fr6}
f(r)=1-\frac{2G_p M}{r}-\frac{r^2}{l_p^2},
\end{equation}
where $G_p=(16\pi p^2g_1)^{-1}$ is a gravitational coupling which is varying along with the running of the cutoff scale and  the  integral constant  $M$ is identified with the physical mass of the black hole. $l_p$ denoting the radius of the asymptotic (A)dS space is also cutoff dependent,
\begin{equation}\label{eq:lp}
l_p^2\simeq -\frac{3g_1}{g_0 p^2}\left[1+\sqrt{1-\frac{g_0}{3g_1^2}(12g_{2a}+3g_{2b}+2g_{2c})}\right].
\end{equation}
From Eqs.(\ref{eq:g_1}) and (\ref{eq:lp}), it is easy to find that solution (\ref{eq:fr6})  is consistent with the usual vacuum solution of GR in the $p\rightarrow 0$ limit.  However, when the energy scale flows to the UV limit, significant difference will appear.
In order to implement quantum corrections to the running coefficients appearing in the classical geometry, we have to determine the relationship between the momentum cutoff $p$ and the radial coordinate $r$.  Considering the trace of generalized Einstein equations (\ref{eq:Einstein}) gives
 \begin{equation}\label{eq:trace}
  \tilde{G}\equiv g_{\mu\nu}\tilde{G}^{\mu\nu}=2p^4g_0+p^2g_1R-2(3g_{2a}+g_{2b}+g_{2c})\Box R=0.
 \end{equation}
From Eq.(\ref{eq:trace}), at small radial distances compared with the Plank scale, it is found that
\begin{equation}\label{eq:UVp}
  p(r)\simeq 2.66 M^{1/4}\lambda_0^{-1/8}r^{-3/4}
\end{equation}
for a fiducial black hole.
On the other hand, when the momentum cufoff flows the IR limit, the high-derivative terms are suppressed automatically, then becomes negligible in Eq.(\ref{eq:trace}). Thus, the approximate relation  $p \sim 1/r$ is obtained. In fact, this identification of infrared cutoff has been discussed rigorously \cite{Bonanno2000}. A numerical analysis of the behaviour of the momentum cutoff with respect to distance is provided in Fig.1 of Ref.\cite{Cai2010}.

Therefore, substituting the approximate relation $p \sim 1/r$ into the black hole solution will yield
\begin{equation}\label{eq:fr}
f(r)\simeq 1-{\eta} \Lambda  G_N-\frac{\Lambda  r^2}{3}-\frac{2 G_N M r}{r^2+G_N\xi}.
\end{equation}
The metric (\ref{eq:metric}) with $f(r)$ given by Eq.(\ref{eq:fr}) denotes a quantum corrected black hole in a (A)dS space-time with a deficit angle ${\eta}\Lambda G_N$.   If the effects of quantum gravity is neglected, the above metric will go back to the metric for classic static black hole in (A)dS spacetime. It is worth pointing that, according to the observations in cosmology, the value of the product  $\Lambda G_N$ is extremely small.  That is, the effect of the deficit angle is negligible for most cases. Furthermore, our purpose is to investigate the quantum corrections to the quasinormal modes of classic Schwarzschild black hole (SBH) solution, so we assume  the terms containing $\Lambda$ in Eq.(\ref{eq:fr}) can be neglected, which means the event horizon is much less than the cosmological horizon (de Sitter radius). Therefore, for our purpose, we take
\begin{equation}\label{eq:fr1}
f(r)= 1-\frac{2 M r}{r^2+\xi},
\end{equation}
where  the Newton's gravitational constant $G_N=1$ is set for simplicity\footnote{Note that since we here take $\hbar=c=G_N=1$, the mass unit is the standard Planck mass $M_p=1/\sqrt{G_N}=1$.}. Clearly, there exist two horizons
\begin{equation}
r_{\pm}=M\pm \sqrt{M^2-\xi},
\end{equation}
when $M^2>\xi$. The inner one is a Cauchy horizon and the outer one is an event horizon.  In the case that $M^2=\xi$, the two horizons are merged and equal to $M$ which is half of the usual Schwarzschild radius. When $M^2<\xi$ the black hole vanishes, corresponding to a naked singularity. It should be stressed that the metric function (\ref{eq:fr1}) is valid providing only in the infrared limit of the AS theory for which the approximate relation $p \sim 1/r$ holds. In order to estimate the range of validity of this approximation, we need to know the approximate form of the metric function $f(r)$ in the UV limit.  In the limit of $p\rightarrow \infty$, from Eq.(\ref{eq:g_1}), $p^2g_1\simeq \xi p^2/16\pi$. Inserting this expression and Eq.(\ref{eq:UVp}) into Eq.(\ref{eq:fr6}) and neglecting the $l_p$ term, we obtain $f(r)\simeq 1- 0.283\xi^{-1}\lambda_0^{1/4}(M r)^{1/2}$, and then the horizon in the UV limit $r_{UV}\simeq 12.5\xi^2\lambda_0^{-1/2}M^{-1}$. When the approximation  $p \sim 1/r$ is valid, we should have $r_{+}>r_{UV}$, equivalently,
\begin{equation}\label{eq:mc}
  M>M_c \simeq \frac{25 \xi ^{3/2}\lambda _0^{-1/2}}{2\sqrt{25 \xi{\lambda _0}^{-1/2}-1}}.
\end{equation}
When $\xi=0.72$ and $\lambda_0=1$, $M_c\approx 1.85 M_p$. Clearly, we should enforce the parameter $\lambda_0 < 625\xi^2$ in order to let the critical mass $M_c$ make sense. Interestingly, from Eq.(\ref{eq:mc}), it is easy to find that $M_c$ is always greater than $\sqrt{\xi}$, this means the naked singularity will never appear.

\section{The quasinormal modes}\label{sec:QNM1}
In the following, we consider the massless scalar perturbations to the black hole.
The propagation of a minimally coupled massless scalar field is described by the Klein-Gordon equation
\begin{equation}
\Box_g \Phi=0.
\end{equation}
Introducing a new radial coordinate (tortoise coordinate) $r_\ast$, satisfying
\begin{equation}\label{eq:tortoise}
d r_\ast=\frac{d r}{f(r)},
\end{equation}
which can be integrated explicitly as
\begin{eqnarray}
r_{\ast}&=&r+\frac{r_{+}(r_{+}+r_{-})}{r_{+}-r_{-}}\ln(r-r_+)
-\frac{r_{-}(r_{+}+r_{-})}{r_{+}-r_{-}}\ln(r-r_-),
\end{eqnarray}
for the metric Eq.(\ref{eq:metric}) with $f(r)$ given by Eq.(\ref{eq:fr1}).
 It is easy to find that when $r$ approaches to event horizon, $r_\ast\rightarrow -\infty$ and when $r\rightarrow \infty$, $r_\ast\rightarrow \infty$, as we expect.
Then if we consider the modes
 \begin{equation}
 \Phi(t,r,\theta,\phi)=\frac{1}{r}\psi(r)Y_{lm}(\theta,\phi)e^{-iwt},
 \end{equation}
 where $Y_{lm}(\theta,\phi)$ are the usual spherical harmonic functions,
 the following standard form of a wave equation is obtained
\begin{equation}\label{eq:RW}
\left[\frac{d^2}{dr_{\ast}^2}+(\omega^2-V(r))\right]\psi(r_\ast)=0,
\end{equation}
with the effective potential
\begin{equation}\label{eq:V}
V(r)=f(r)\left(\frac{l(l+1)}{r^2}+\frac{f'(r)}{r}\right),
\end{equation}
where $l$ is the multipole quantum number which arises from the separation of angular variables by expansion into spherical harmonics.
Quasinormal modes are solutions of the wave equation (\ref{eq:RW}), satisfying the boundary conditions that  the wave functions represent   pure ingoing waves at the event horizon
\begin{equation}
\psi\sim e^{-i\omega r_\ast}, \;\;\; r_\ast\rightarrow-\infty
\end{equation}
and  pure outgoing ones  at spatial infinity
\begin{equation}
\psi\sim e^{i\omega r_\ast}, \;\;\; r_\ast\rightarrow +\infty.
\end{equation}
Because of the complexity of the effective potential (\ref{eq:V}), it is generally difficult to obtain an exact analytic solution for wave equation (\ref{eq:RW}). Therefore,  some approximation is usually used to solve the equation.

By using the method suggested by Mashhoon\cite{Mashhoon1984}, in which the potential $V(r(r_\ast))$   is modeled approximately by the P$\ddot{\mathrm{o}}$schl-Teller potential, the quasinormal modes $\omega$ is obtained
\begin{equation}
\omega=\sqrt{V_0-\frac{1}{4}\alpha^2}-i\left(n+\frac{1}{2}\right)\alpha,\;\;\;\; n=0,1,2,\cdots
\end{equation}
Here the parameter $\alpha=(-V_2/2V_0)^{1/2}$ where $V_0$ and $V_2$ are the height and curvature  of the effective potential at its maximum, respectively. This method gives quite accurate results for the regime of high multipole numbers $l\rightarrow \infty$. In this limit,  the maximum of the potential for a heavy black hole($M^2\gg \xi$) is located at
\begin{equation}
r_{\mathrm{max}}\simeq 3M-\frac{5\xi}{9M}.
\end{equation}
Then, we obtain that $\omega=\omega_R-i\omega_I$, where
\begin{eqnarray}\label{wRwI}
&&\omega_R \simeq\frac{1}{3\sqrt{3}M} \left(l+\frac{1}{2}\right)\left(1+\frac{\xi}{9M^2}\right),\\
&&\omega_I \simeq \frac{1}{3\sqrt{3}M} \left(n+\frac{1}{2}\right)\left(1-\frac{2\xi}{27M^2}\right).
\end{eqnarray}
It is obvious that the quantum corrections increase  the real part of quasinormal frequencies $\omega_R$, while decrease the imaginary part $\omega_I$. When $\xi>27M^2/2$, $\omega_I$ becomes negative. This seems to  mean an instability of the black hole. However, this actually  can not happen, because the mass of the black hole has a low limit $M_c>\sqrt{\xi}$. It should be emphasized that the above results are not justified for an extremely small black hole which $M\sim M_c$. Furthermore, it is  pointed out that these modifications to quasinormal frequencies are not the same as those suggested by the string theory. For example, the imaginary part  QNMs of scalar type  for high dimensional Gauss-Bonnet black holes is decreasing when the Gauss-Bonnet coupling parameter is increasing, while the real part takes on a complicated behavior \cite{Konoplya2008}.

In order to  test the results obtained above, we here use the Frobenius method, which was first introduced to QNM calculation by Leaver\cite{Leaver1985}, to perform a more accurate numerical
analysis of the evolution of perturbations for AS black holes. To match the boundary conditions of QNM modes at event horizon ($r=r_+$) and infinity ($r=\infty$), we expand the solution of (\ref{eq:RW}) as the following series
 \begin{equation}\label{series}
\psi(r)=e^{-\rho r}(r-r_{-})^{-\rho (r_{+}+r_{-})}z^{\rho  \frac{r_+\left(r_++r_-\right)}{\left(r_+-r_-\right)}}\sum^{\infty}_{n=0}a_n z^n,
 \end{equation}
where $z=(r-r_+)/(r-r_-)$, $\rho=-i \omega$ and $a_0$ is taken to be $a_0=1$.
Substituting the series (\ref{series}) into Eq.(\ref{eq:RW}), we obtain a seven-term recurrence relation for the expansion coefficients $a_n$ in Eq.(\ref{series}):
\begin{equation}
\sum_{i=0}^{\min(6,n)}c_{i,n}a_{n-i}=0\;\;\; \mathrm{for} \;\;\;n>0,
\end{equation}
where
\begin{eqnarray}
c_{0,n}&=&n \left(n+\frac{2 \rho  r_+ \left(r_-+r_+\right)}{r_+-r_-}\right),\nonumber\\
c_{1,n}&=&2 n-2 n^2 -1-l (1+l)-\frac{4 (n-2) (n-1) r_-}{r_++r_-}\nonumber\\
&-&\frac{(1+l (1+l)+2 (n-2) n) r_-}{r_+}+\frac{8 \rho ^2 r_+^2 \left(r_-+r_+\right)}{r_--r_+}\nonumber\\
&+&\frac{4 \rho  \left((n-1) r_-^2+(3 n-4) r_- r_++(2 n-1) r_+^2\right)}{r_--r_+},\nonumber\\
c_{2,n}&=&(n-1)^2+\frac{(n-3) (n-1) r_-^2}{r_+^2}-\frac{4 \rho ^2 r_+ \left(5 r_-^2+14 r_- r_++r_+^2\right)}{r_--r_+}\nonumber\\
&+&\frac{2 \rho  \left((n-2) r_-^4+2 (5n-13) r_-^3 r_+\right)}{r_+^3-r_-^2 r_+}\nonumber\\
&+&\frac{2 \rho  \left((19n-40) r_-^2 r_+^2+2 (14n-25) r_- r_+^3+2 (n-1) r_+^4\right)}{r_+^3-r_-^2 r_+}\nonumber\\
&+&r_- \left(\frac{64+8 l (1+l)-58 n+13 n^2}{r_+}+\frac{8 (n-2)}{r_-+r_+}\right),\nonumber\\
c_{3,n}&=&2 r_- \left(2 \rho  \left(21-8 n-36 \rho  r_-\right)-16 \rho ^2 r_+\right.\nonumber\\
&-&\frac{40 \rho  r_- \left(3-n-2 \rho  r_-\right)}{r_--r_+}-\frac{\left(l+l^2+(n-3) (4n-15)\right) r_-}{r_+^2}\nonumber\\
&+&\frac{45 n-8 n^2+2 (2n-9) \rho  r_- -9 \left(7+l+l^2\right)}{r_+}\nonumber\\
&+&\left. \frac{4 \left(2 l (1+l)+(n-3)^2-4 (n-3) \rho  r_-\right)}{r_-+r_+}\right),\nonumber\\
c_{4,n}&=& \frac{r_-}{r_-^2 r_+^3-r_+^5} \left(r_-^4 \left((n-5)^2+4 \rho  r_+ \left(1-\rho  r_+\right)\right)\right.\nonumber\\
&-&r_+^4 \left(n-5+2 \rho  r_+\right) \left(n-3+2 \rho  r_+\right)\nonumber\\
&-&\left.r_- r_+^3 \left(8 l (1+l)+(n-4) (13n-46)-2 \rho  r_+ \left(110-29 n-32 \rho  r_+\right)\right)\right.\nonumber\\
&-&\left. 2 r_-^2 r_+^2 \left(21-5 n-2 \rho  r_+ \left(71-18 n-38 \rho  r_+\right)\right)\right.\nonumber\\
&+&\left.r_-^3 r_+ \left(8 l (1+l)+(n-4) (13n-54)+2 \rho  r_+ \left(7n-22-48 \rho  r_+\right)\right)\right),\nonumber\\
c_{5,n}&=&\frac{r_-^2}{r_+^3} \left(\frac{8 \rho ^2 r_+^2 \left(r_-+r_+\right) \left(4 r_-+r_+\right)}{r_--r_+}\right.\nonumber\\
&-&\frac{4 \rho  r_+ \left((3n-14) r_-^2+(34-7 n) r_- r_+-2 (n-5) r_+^2\right)}{r_--r_+}\nonumber\\
&-&\frac{\left(61+l+l^2+2 (n-11) n\right) r_-^2+2 \left(l+l^2+(n-5) (4n-19)\right) r_- r_+}{r_-+r_+}\nonumber\\
&-&\left.\frac{\left(49+l+l^2+2 (n-10) n\right) r_+^2}{r_-+r_+}\right),\nonumber\\
c_{6,n}&=&\frac{r_-^3 \left(n-6+2 \rho  \left(r_-+r_+\right)\right) \left(r_+ \left(6-n-2 \rho  r_+\right)+r_- \left(n-6-2 \rho  r_+\right)\right)}{\left(r_--r_+\right) r_+^3}.
\end{eqnarray}
It is easy to find that when there is no effect of quantum gravity(i.e. $r_-=0$), the recurrence relations above become  three-term relations, and if we further set $r_+=1$, the recurrence relations are reduced to be Eqs.(6)-(8) for scalar perturbations in Ref.{\cite{Leaver1985}}.
By using gaussian elimination method for several times \cite{Leaver1990,Cardoso2003}, we can reduce the seven-term recurrence relation to
a three-term relation, which has the form
\begin{eqnarray}
\alpha_1 a_1+\beta_1 a_0&=&0\\
\alpha_n a_n+\beta_n a_{n-1}+\gamma_n a_{n-2}&=&0\;\; \mathrm{for}\;\; n=2,3,\cdots
\end{eqnarray}
Using the coefficients $\alpha_n$, $\beta_n$ and $\gamma_n$, according to Leaver\cite{Leaver1985}, QNM frequencies are given by the vanishing point of the following continued fraction equation
\begin{equation}
0=\beta_1-\frac{\alpha_1\gamma_2}{\beta_2-}\frac{\alpha_2\gamma_3}{\beta_3-}\frac{\alpha_3\gamma_4}{\beta_4-}\cdots.
\end{equation}
For practice, a finite but very large $n$ is chosen to solve the algebraic equation for the required accuracy.

We list the numerical values of the frequencies of fundamental and first-overtone QNMs obtained by using above continued fraction method for different low values of $l$ in table \ref{tab:1}, where we have set the mass of the black hole $M=1$ and chosen $n=150$. It is found without surprise that $\omega_R$ for black holes with $\xi$-correction are greater than those for Schwarzschild  black holes ($\xi=0$), meanwhile, the value of $\omega_I$ is decreased. This result is consistent with the above analysis for the eikonal limit case ($l\rightarrow \infty$). Here we note that the sizable effects displayed in table \ref{tab:1} refer to a mini black hole. It goes without saying that, for an astrophysical black hole, the difference between the numbers for black holes with and without $\xi$-correction  would be very small.
\begin{table}
  \centering
\begin{tabular}{c|c|c|c|c}
\hline\hline $l$ & $n$ & $\omega(\xi=0.64)$& $\omega (\xi=0.36)$ & $\omega(\xi=0)$  \\
\hline
0 & $0$ & $0.121296 -0.0943462 i$ &$ 0.116741-0.100288 i$  & $0.110455 -0.104896 i$   \\
 & $1$  & $0.0965272 -0.303120 i$ &$0.094949-0.328204 i$ & $0.0861803 -0.348308 i$   \\
\hline
1 & $0$ & $0.319953 -0.0893338 i$&$0.306755 - 0.094030 i$ & $0.292936 -0.097660 i$   \\
 & $1$  & $0.299239 -0.275377 i$&$ 0.283163 - 0.292448 i$  & $0.264449 -0.306257 i$   \\
\hline
2 & $0$ & $0.527530 -0.0887348 i$&$0.505797 -0.093285 i$ & $0.483644 -0.0967588 i$   \\
 & $1$  & $0.513670 -0.269172 i$&$0.489538-0.284035i$  & $0.463851 -0.295604 i$   \\
\hline
3 & $0$ & $0.736347 -0.0885638 i$&$0.706043-0.093076i $ & $0.675366 -0.0964996 i$   \\
  & $1$ & $0.726141 -0.267251 i$&$0.693984-0.281421 i$  & $0.660671 -0.292285 i$   \\
\hline \hline
\end{tabular}
\caption{The frequencies of fundamental and first-overtone QNMs  for some different low values of $l$ of black hole with quantum corrections ($\xi=0.64, 0.36$) and usual SBH ($\xi=0$), by using the continued fraction method. Here the mass of the black hole $M$ is set to be $1$.}
\label{tab:1}
\end{table}

For a more intuitive investigation of quantum-gravity corrections to perturbations of black hole, we would like to see the object picture of the quasinormal oscillations.  For this purpose,  without implying the stationary ansatz $(\Phi \varpropto e^{-i\omega t})$,  we introduce the null coordinates $u = t - r_{\ast}$ and $v = t + r_{\ast}$. Then,  the wave equation Eq.(\ref{eq:RW}) is modified to be
\begin{equation}\label{eq:uv}
4\frac{\partial^{2}\psi}{\partial u \partial v}+V(r_{\ast})\psi=0.
\end{equation}
The above equation can be numerically integrated by the ordinary finite
element method. Using the Taylor expansion, one find \cite{Gundlach1994}
\begin{eqnarray}
\psi_{N}&=&\psi_{E}+\psi_{W}-\psi_{S}
 -\frac{\Delta^2}{8}V\left(\frac{v-u}{2}\right)(\psi_{W}+\psi_{E})
+\mathcal{O}(\Delta^{4}),
\end{eqnarray}
where $N$, $W$, $E$ and $S$ are the points of a unit grid on the
$u-v$ plane which correspond respectively to the points ($u+\Delta$, $v+\Delta$),
($u+\Delta$, $v$), ($u$, $v+\Delta$) and ($u$, $v$).  Here $\Delta$ is
the step length of the change of $u$ or $v$. Because the QNMs of black holes are
insensitive to the initial conditions, we begin with a Gaussian
pulse of width $\sigma$ centered on $v_{c}$ at $u = u_{0}$ and set
the wave function to zero at $v = v_{0}$  \cite{xi2005}.

\begin{figure}
\centering
\includegraphics[width=\linewidth]{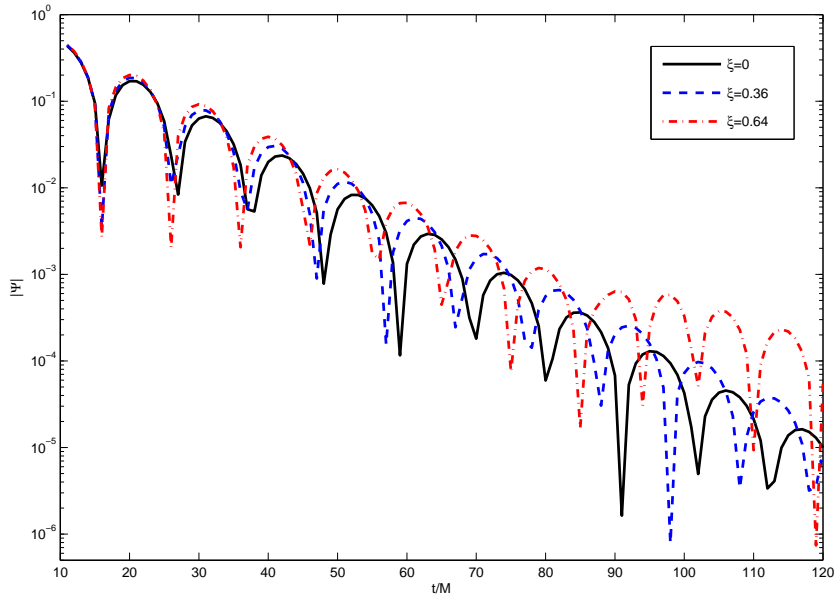}
\caption{The time-domain profiles of massless scalar perturbations for the SBH($\xi=0$) and the AS black holes in the IR limit with the same mass and different values of parameter $\xi$. Here multipole quantum number $l=1$. }
\label{fig:1}
\end{figure}

 As is shown that the time evolution of the perturbations follows the usual dynamics. After a transient stage heavily dependent on the initial conditions, there is an  exponential damping of the perturbations, which can be characterized by QNMs, and what follows is a  power-low tails at late time. In Fig.\ref{fig:1},  the time-domain profiles of massless scalar perturbations for the SBH($\xi=0$) and the AS black holes with the same mass and different value of parameter $\xi$ are plotted logarithmically. It can be observed that the oscillating quasi-period and the damping time scale  of perturbations for SBH are longer and smaller  than those for AS black holes, respectively. This is reasonable,  because   the greater the value of $\omega_R$ takes, the faster the perturbation oscillates and the less the value of $\omega_I$ acquires, the faster the perturbation damps and,  as we have found before, the $\xi$-corrections can make the real part of QNMs increase and the imaginary part decrease.

\section{Conclusions}
\label{sec:conclusion}
In conclusion,  we find that due to the gravitational nonperturbative effects,  the static, spherically symmetric  solution in the infrared limit of  asymptotic safe gravity can denote a quantum corrected black hole in (A)dS space-time with a deficit angle. The effect of the deficit angle is negligibly small inferred from cosmological observations and, for astrophysical black hole, the corrections yielded by the running of gravitational couplings are extremely small. However, the effect of quantum corrections is still possibly significant for miniature black holes. Of course, if the black hole is extremely small, the metric function for the spacetime outside to black hole we discussed above will not hold. A concise estimation for the valid condition has been given.  By investigating the effects of this quantum correction to the quasinormal modes of a test scalar field on the AS black hole background in both analytical and numerical approaches,  we show that  the quasi-period frequencies and the damping rate of oscillations are respectively enhanced and weakened by the quantum correction, but the stability of black hole solution keeps unchanged. The quantum effects on the QNMs for black holes in asymptotic safe gravity is distinctive from those in other scenarios of quantum gravity.

It should be pointed out that,  just as Reissner-Nordstr\"{o}m (RN) black hole, this kind of black hole solution has also two horizons and when the inner Cauchy horizon approaches to zero, both of them go back to usual Schwarzschild black hole. However, they are two fundamentally different black hole solutions. Although there is little difference between the external metric far away from these two kind of black hole,  for the locations near the event horizon, the difference between the external metric of these two kind of black hole is remarkable and this results in the difference of their QNMs. It is noticed that in the calculations of QNM frequencies for RN black holes using Frobenius series method, one can obtain directly a three-term recurrence relations among the coefficients\cite{Ohashi2004}.

The quasinormal modes for  black holes in anti-de Sitter (AdS) spacetime may have the
AdS/CFT holographic interpretation.  The highly damped QNMs of a quantum corrected black hole\cite{Babb2011}, in which the importance of the existence of two distance scales in the metric, is discussed. It would seem that the discussion will get more complicated and potentially more interesting when there is a position dependent momentum cut-off in the asymptotic safe scenario of gravity. Therefore, it is very interesting to investigate the the asymptotically high overtones of the QNMs for AS black holes in AdS spacetime, which is beyond the present study and we hope that this point will be clarified in a future work.

\section*{Acknowledgments}
We would like to thank Xi-Chen Ao for valuable discussions. This work is supported by  Innovation Program of Shanghai Municipal Education Commission under Grant No. 09YZ148 and SRFDP under Grant No. 200931271104.


\end{document}